# Highly effective and isotropic pinning in epitaxial Fe(Se,Te) thin films grown on CaF$_2$ substrates

V. Braccini,[1,a)] S. Kawale,[1,2] E. Reich,[3] E. Bellingeri,[1] L. Pellegrino,[1] A. Sala,[1,2] M. Putti,[1,2] K. Higashikawa,[4] T. Kiss[4], B. Holzapfel,[3] and C. Ferdeghini[1]

[1] CNR – SPIN Genova, Corso Perrone 24, 16152 Genova, Italy

[2] DIFI, University of Genova, Via Dodecaneso 33, 16145 Genova, Italy

[3] IFW Dresden, Helmolzstr. 20, 01069 Dresden, Germany

[4] Kyushu University, 744 Motooka, Nishi-ku, Fukuoka 819-0395

## Abstract

We report on the isotropic pinning obtained in epitaxial Fe(Se,Te) thin films grown on CaF$_2$(001) substrate. High critical current density values - larger than 1 MA/cm$^2$ in self field and liquid helium – are reached together with a very weak dependence on the magnetic field and a complete isotropy. Analysis through Transmission Electron Microscopy evidences the presence of defects looking like lattice disorder at a very small scale, between 5 and 20 nm, which are thought to be responsible for such isotropic behavior in contrast to what observed on SrTiO$_3$, where defects parallel to the $c$-axis enhance pinning in that direction.

---

[a)] Author to whom correspondence should be addressed. Electronic mail: valeria.braccini@spin.cnr.it

Superconducting iron chalcogenides, also called 11 family, are the simplest among iron-based superconductors. They are quite attractive being of easy fabrication and relatively safe synthesis thanks to the absence of arsenic, they show upper critical fields approaching 50 T,[1] and their critical temperature $T_c$ can be enhanced from the relatively low 15 K of the bulk up to 21 K in thin films thanks to substrate-induced compressive strain.[2] The main drawback of the chalcogenides was their relatively low critical current density $J_c$: the recent report of values exceeding 1 MA/cm$^2$ in self-field on 11 thin films with a $CeO_2$ buffer layer epitaxially grown on single-crystalline and coated conductors substrates[3] make chalcogenides very promising for high-field applications at low temperature. In this paper authors hypothesize that pinning defect could arise from the adopted $CeO_2$ buffer layer.

In our previous publications,[2,4] we have studied $FeSe_{0.5}Te_{0.5}$ thin films grown by Pulsed Laser Deposition (PLD) on different oxide and fluoride substrates, evidencing the major effect the substrate can produce on the strain and therefore on $T_c$, $J_c$, its anisotropy, and the pinning mechanisms. Recently, Fe(Se,Te),[5,6,7,2] Co-doped $BaFe_2As_2$[8] as well as $RE$FeAs(O,F) ($RE$: Nd and Sm)[9] thin films have been deposited on $CaF_2$ and showed very high crystalline quality and excellent superconducting properties due to a growth free of oxygen contamination, and thanks to the introduction of a moderate in-plane compressive strain.[7]

In this Letter we focus on superconducting and microstructural properties of a thin film grown on $CaF_2$ with a $T_c$ of 20 K which exhibits $J_c$ values as high as 1.1 MA/cm$^2$ at 4 K in self-field – consistent with the highest reported for the 11 phase[3] - with a very weak magnetic field dependence and an almost complete isotropy. The behavior of $J_c$ and the isotropic pinning is discussed with the support of Transmission Electron Microscopy (TEM).

The thin film analyzed in this paper is 180 nm thick and was grown by PLD on $CaF_2$*(001)* single crystal at 550°C, with 3 Hz (248 nm wavelength) as laser repetition rate, 2 J/cm$^2$ as laser fluency and 5 cm as target – substrate distance.[2] Hall bar shaped microbridges of 10, 20 and 50 μm of width and 65 μm distance between voltage contacts were realized by conventional photolithography and water-cooled dry etching (Ar ions, 500 eV, 0.1 mA/cm$^2$). $T_c$ measured on different filaments is the same, showing an onset of 20.6 K and a $T_{c,0}$ value of 19 K, the highest ever reported on films grown on $CaF_2$ substrate, quite similar to the value measured on films with $CeO_2$ buffer layer[3] and almost as high as the 21 K we reported on $LaAlO_3$.[10] Transport critical current measurements of the microbridges as a function of temperature, magnetic field and angle were performed in a 9 T Physical Properties Measuring System (PPMS) operating with a sample rotator. Current vs. Voltage characteristics (I-V) were acquired sweeping the current with exponentially increasing steps, with the aim to avoid heating problems. The critical current is defined with the 1 µV/cm criterion. In-plane distribution of local $J_c$ in the film was visualized by low temperature scanning Hall-probe microscopy (LT-SHPM). Transmission electron microscopy (TEM) was performed using a FEI Tecnai T20 ($LaB_6$, 200 kV) and an FEI Titan 80–300, operating at 300 kV (FEG) with an image $C_s$ corrector. The lamellae were prepared with the in-situ lift-out method in focused ion beam (FIB) device.[11]

XRD patterns showing *(00l)* reflections only confirms very high crystalline quality of the film and the optimum *c*-axis alignment of the growth; the *ω* scan of the *(001)* reflection shows a sharp full width at half maximum (FWHM) of 0.8°, while the *φ* scan of the *(101)* reflection only reveals sharp reflections at every 90° - with a FWHM of ≈ 0.8° - indicating a single in-plane epitaxy of the growth.[4] The *a* cell parameter of 3.729 Å corresponds to the high 20.6 K $T_c$.[10] $CaF_2$ has the same effect on the growth than the oxide substrates, being the correlation between increase in $T_c$ and shrinkage in the *a*-axis unambiguously driven by the compressive strain.

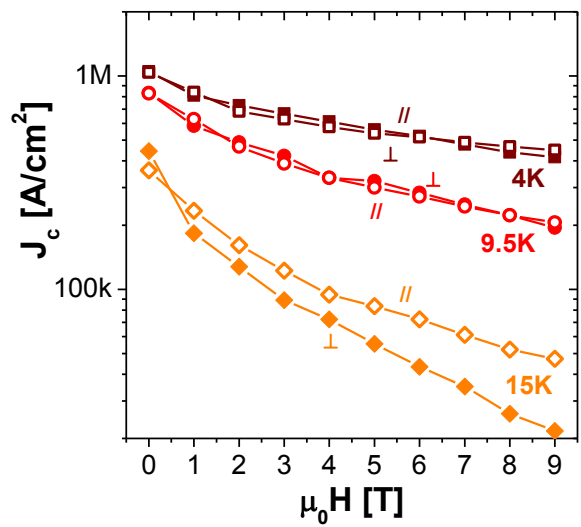

FIG.1: $J_c$ (*H*) at 4, 9.5 and 15 K for the film deposited on $CaF_2$, for *H//ab* (hollow symbols) and *H⊥ab* (full symbols).

$J_c$ measured on the 10 μm wide bridge as a function of the magnetic field in two directions are reported at of 4, 9.5 and 15 K in Fig. 1. In self-field $J_c$ is above 1 MA/cm$^2$ at 4 K, which is as high as the outstanding value reported on films grown with a CeO$_2$ buffer.[3] At $T$ = 15 K, which is 80% of $T_{c,0}$, we measured $J_c$ (0T, $t$=0.8) = 0.4 MA/cm$^2$ instead of the 0.2 MA/cm$^2$ recorded with the CeO$_2$ buffer,[3] while in thin films grown on CaF$_2$ for the same reduced $t$ = 0.8 – corresponding to $T$ = 12 K - $J_c$ of 0.05 MA/cm$^2$ was previously reported.[7]

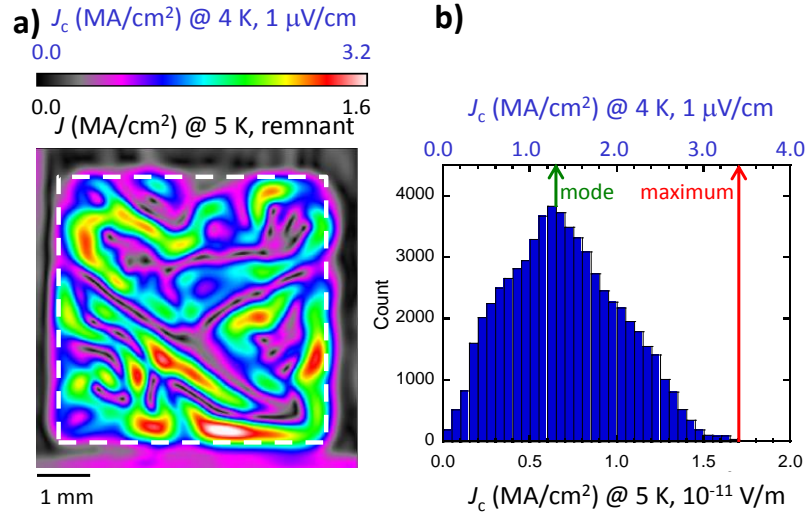

FIG.2: a) In-plane distribution of current density in the film visualized by low temperature scanning Hall-probe microscopy (LT-SHPM) at 5 K in a remnant stage. The distribution corresponds to that of $J_c$ at 5 K and 10$^{-11}$ V/m in electric field criterion. Considering the temperature and the electric field dependence, $J_c$ at 4 K and 1 μV/cm is also estimated on the second axis. The broken line indicates the dimension of the film. b) Histogram of local $J_c$ distribution in the film. The maximum value is almost three times larger than the mode; 3.4 MA/cm$^2$ as the maximum and 1.3 MA/cm$^2$ as the mode were estimated for 4 K and 1 μV/cm.

In order to evaluate the film homogeneity, in Fig. 2 we report the local $J_c$ distribution in an unpatterned film visualized by LT-SHPM. The magnetic field distribution was measured at 5 K in a remnant state, and then the corresponding distribution of current density was estimated by solving an inverse problem of the Biot-Savart law.[12] According to the critical state model, it is possible to regard the distribution of current density shown in Fig. 2 as that of $J_c$ in this film. The validity of this method was already confirmed for REBCO coated conductor.[13] Here the measurement was performed at 5 K and the corresponding electric field criterion was estimated to be 10$^{-11}$ V/m by a relaxation measurement.[14] By considering the dependences of temperature and electric field criterion, $J_c$ at 4 K and 1 μV/cm was also estimated on the second axis. The $J_c$ distribution is rather inhomogeneous: the maximum value was estimated to be higher than 3 MA/cm$^2$ while the mode was around 1 MA/cm$^2$, in good agreement with the value measured by the four-probe method. There are regions where $J_c$ is quite low, and we think that improvement in the overall homogeneity might straightforwardly lead to a transport $J_c$ of 3 MA/cm$^2$. Regions with lower $J_c$ seem to be correlated to particular crystallographic directions: for example, the main defect is parallel to the (*a00*) direction. Very recently, a similar study of $J_c$ in FeSe$_{0.5}$Te$_{0.5}$ thin films deposited on CaF$_2$ estimated by means of LT-SHPM was reported: $J_c$ was estimated 1 MA/cm$^2$, three times higher than the value directly measured by the four-probe method.[15, 16]

Analyzing the behavior of $J_c$ with the field shown in Fig. 1, we notice a very weak field dependence especially at low temperatures – $J_c$ halves increasing the magnetic field from 0 to 9 T – but the more striking feature is the complete isotropy of $J_c$ at 4 and 9.5 K. In Fig. 3a) we report the $J_c$ versus the orientation with respect to the magnetic field at 9 T at 4 and 15 K (corresponding to a reduced temperature $t = T/T_c \approx 0.8$). The angular $J_c$ is compared with the angular behavior reported for the thin film deposited on SrTiO$_3$.[2] At 4 K $J_c$ is isotropic in the whole angular range, while at temperatures closer to $T_c$ a wide peak emerges for $\theta = 90°$, as expected from the intrinsic mass anisotropy of the material, while no evidence of mere *c*-axis correlated pinning is present as for the film grown on SrTiO$_3$, where we observed at all temperatures and fields a pronounced anisotropy and a large *c*-axis peak ($\theta = 0°$) sign of strong correlated *c*-axis pinning.[17] A similar *c*-axis pinning was also observed by others in thin films grown on CaF$_2$,[18] although the *c*-axis correlated peak was very broad and only appeared for fields above 3 T. The majority of measurements on films of the 11 phase[5, 16, 19] simply showed peak for *H // ab*, consistent with the expected intrinsic anisotropy of the phase, as we previously reported for 'clean' films grown on LaAlO$_3$.[2]

In order to investigate the correlation between microstructure and pinning mechanisms, TEM studies were performed on thin films whose angular behavior is reported in Fig. 3a). Fig. 3b) shows the microstructure of the sample deposited on CaF$_2$. In the first panel a clear reaction layer can be observed, as was also seen on Ba-122 films deposited on CaF$_2$ and attributed to a BaF$_2$ formation at the interface,[8] as well as on 11 films grown on CaF$_2$ and found to be part of the substrate[5] or due to fluorine ions incorporated into the Fe(Se,Te) phase through substitution of Se ions or intercalation between the layers.[20] The microstructure of the film does not show any granular structure and large extended defect throughout the whole film; however, small regions appear where the lattice seems to be disturbed on a very local scale (5–20 nm) (Fig. 3b) second panel, marked in black). In the third panel of Fig. 3b) a higher resolution image is shown: at some point the atoms start to deviate from the ideal plane while around the defects the lattice grows fine again. Such defects might originate from a different local stoichiometry between Se and Te - indeed FeTe and FeSe phases show quite different $c$ axis values, 6.285 and 5.486 Å respectively. The tendency to form clusters instead of a uniform distribution was also evidenced at the surface by STM measurements were it is possible to distinguish between the two elements just based on their the different heights.[21] Given their size and distribution it is likely that such kind of defects are the origin of the isotropic pinning present in films grown on $CaF_2$. TEM analysis on a thin film deposited on $SrTiO_3$ is reported in Fig. 3c). The interface between the film and the substrate does not show any reaction layer in strong contrast to the film on $CaF_2$.

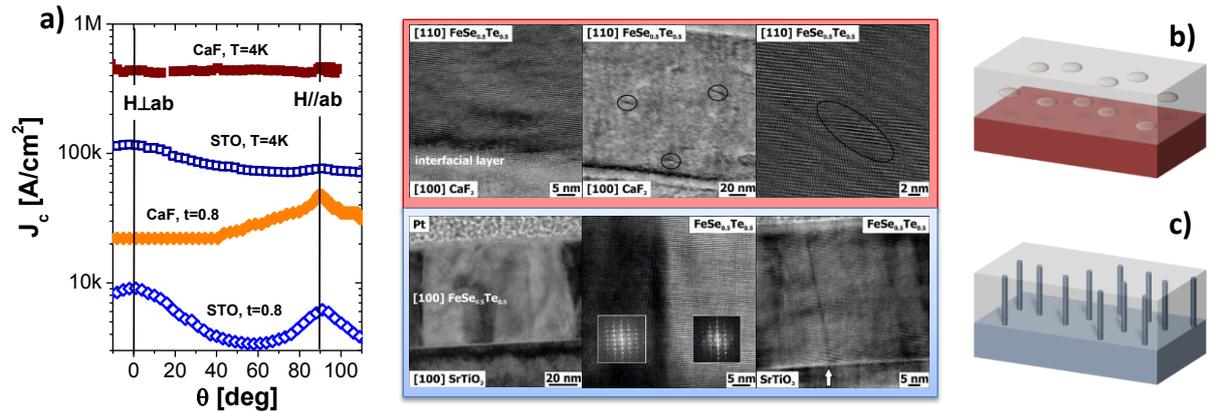

FIG. 3: a) $J_c(\theta)$ at 9 T and 4 K and a reduced temperature $t = T/T_c$ corresponding to about 0.8 for the films deposited on $SrTiO_3$ and $CaF_2$. b) First panel: HRTEM image showing the interfacial layer between the $CaF_2$ substrate and the superconducting thin film. Second panel: regions – marked in black - where the lattice seems to be disturbed on a very local scale. Third panel: HRTEM image showing one defect corresponding to the point where the atoms start to deviate from the lattice lines. Fourth panel: a sketch of the isotropic defects distributed in the film. c): TEM images on thin film grown on $SrTiO_3$. First panel, diffraction contrast reveals grain structure. Second panel, HRTEM image taken at the FEI Titan³ confirms different grain orientations; insets show FFTs from the separate grains. A small in-plane rotation out of the [100] zone axis can be observed. Third panel: HRTEM image showing one single defect parallel to the $c$-axis which extends through the complete film thickness. Fourth panel: a sketch of the nanorods parallel to the $c$-axis.

The overview bright field image ((Fig. 3c), first panel) taken in the [100] zone axis reveals a grain like structure. The high resolution image confirms the grain structure with grains oriented in the [100] zone axis (dark) and bright ones being rotated in plane out of the [100] zone axis by a small angle (FFT insets in Fig. 3c), second panel). Additionally to these low angle grain boundaries which can be found throughout the complete lamella, single defects parallel to the *c*-axis can be seen (Fig. 3c), third panel) which are compatible with the network of columnar defects parallel to the *c*-axis shown by STM previously hypothesized[4] to contribute, together with the small angle grain boundaries, to enhance the pinning parallel to the *c*-axis recorded by the highest $J_c$.[2] Such small angle grain boundaries were not observed in the film deposited on $CaF_2$: a lower in-plane epitaxy was also found from the $\varphi$ scan of the *(101)* reflection performed on the film on $SrTiO_3$, which showed a FWHM of about 1.4°, higher with respect to the 0.8° measured on the film deposited on $CaF_2$.

The *c* axis correlated pinning observed for films deposited on $SrTiO_3$ is similar to what reported for 122 films,[22] where for all magnetic fields, $J_c$ of the thin film grown on *001*-oriented $(La,Sr)(Al,Ta)O_3$ substrate using an intermediate template of 100 unit cells of $SrTiO_3$ exhibits a strong and broad *c*-axis peak. At low fields pinning becomes effective over all $\theta$, significantly flattening $J_c(\theta)$. At larger fields, the *c*-axis peak still remains pronounced but in a smaller range around 0° so the *ab* plane peak due to the mass anisotropy emerges. The microstructural origin of strong vortex pinning was revealed by TEM images, which showed a high density of randomly distributed columnar defects with average diameter of 5 nm. The pinning in Co-doped Ba122 thin films was even tuned by combining self-assembled nanorods, which act as strong *c*-axis correlated pins enhancing $J_c$ along the *c*-axis, and artificially introduced nanoparticles which are effective over a wide angular range and provide additional *ab*-plane pinning, generating a precipitate landscape that develops an almost isotropic pinning.[23] With our work we show that it is instead possible to tune the pinning from *c*-axis, to isotropic to *ab*-plane by changing substrate and/or inducing small stoichiometry variations between Se and Te in order to control size and density of the isotropic defects induced by the use of the $CaF_2$ substrate.

**ACKNOWLEDGEMENTS**

This work has been partially supported by FP7 European project SUPER-IRON (No. 283204).